\documentclass[twocolumn,showpacs,showkeys,preprintnumbers,amsmath,amssymb]{revtex4}

\usepackage{graphicx}
\usepackage{dcolumn}
\usepackage{bm}

\begin{document}


\title{Evolution of Metal Structure at Intense Plastic Strains:
Molecular Dynamics Simulation}

\author{L.S.Metlov}
\email{metlov@mail.donbass.com}
\affiliation{Donetsk Physico-Technical Institute, Ukrainian
Academy of Sciences,
\\83114, R.Luxemburg str. 72, Donetsk, Ukraine}

\thanks{This work was supported by the Fund of Budget Researches of the National
Academy of Ukraine (topic 108) and by Donetsk Innovation Center
(provision of computer and internet).}

\date{\today}

\begin{abstract}

The kinetics of dislocations is studied with computer simulation
at loadings of different intensity. It is established that the
dislocations have a few different structural states. The
dislocations "with the micropore" play important role in the
formation of large curved boundaries, and, as a consequence, in
the formation of fine grains. Alternation of elastic and inelastic
strain stages is established too. At shear loading, in view of
special kinetics, the system would have to accumulate the whole
set of dislocations leading to the formation of new boundaries and
fine grains.
\end{abstract}


\maketitle

\section{Introduction}

Processing of materials by using the intense plastic strains (IPS)
to obtain the grains with the size of about tens and hundreds
nanometers is a promising direction of modern technologies
\cite{PMVM02}. During the IPS the processes with high strain rate
take place. As a consequence, there is not enough time to
establish the equilibrium state in the systems, and the process
has essentially the non-equilibrium character. From the
thermodynamic point of view, fast strain of a metal sample is
similar to the process of quenching with the difference, that the
transition is realised not by a change in temperature, but with a
change of pressure. Owing to this, at the IPS the energy pumping
into the sample occurs through the creation of structural
non-equilibrium objects, such as fine grains, additional
non-equilibrium strongly curved boundaries, etc. It is possible to
obtain many qualitative results about features of the IPS with the
molecular dynamics simulation on rather small systems of about
1000 atoms and less.

The computer experiments, the results of which are discussed
below, are carried out using the pair Lennard-Jones potential of
interparticle interaction in the form:
  \begin{equation}\label{a1}
    U_{ijlk}=E_{b}((\frac{r_{0}}{r_{ijlk}})^{12}-2(\frac{r_{0}}{r_{ijlk}})^{6}),
  \end{equation}
where $r_{ijlk}=\sqrt{(X_{ij}-X_{lk})^{2}-(Y_{ij}-Y_{lk})^{2}}$ -
is the distance between the particles numbered $i, j$ and $l, k$
with Cartesian coordinates $X_{ij}, Y_{ij}$ and $X_{lk}, Y_{lk}$.
The indices $i, l$ numerate atoms in the lattice along the
$Y$-direction, $j, k$ do the same along the $X$-direction. $E_{b},
r_{0}$ - are the binding energy and equilibrium interparticle
distance in the two atomic system, respectively.

The regimes of loading the samples, used below in the numerical
experiments, are typical of methods of equalchannel torsion
\cite{PMVM02}, \cite{Vali00}, and for formation of nanostructures
through friction \cite{KEChG01}. Besides, it is established that
nanostructures are formed in the near-surface layer of rails
during their operation. Really, the surfaces of a rail and a wheel
represent some set of roughness' of different scale. In the
process of "rolling", a roughness of a wheel uniformly or
non-uniformly is superimposed over a roughness of a rail. As a
result, each roughness undergoes uniform or non-uniform uniaxial
strain. Besides, the tractive forces developed at movement of a
rolling-stock, create shear forces on the roughness. Qualitatively
the same picture is present during the formation of nanostructures
through friction under pressure \cite{KEChG01}.

\section{Uniaxial compression of sample under rigid boundary conditions}

The rigid boundary conditions can occur in a material when a soft
microscopic inclusion is surrounded with a more rigid matrix. The
rigid boundary conditions are idealistic, but they can be useful
at study of structural relaxation in a small bulk. It is known
that in a small bulk the equilibrium state is established much
faster without the special influence of the other bulk. Therefore
for the time of the establishment of the equilibrium state it is
possible to isolate the investigated microbulk from the other bulk
through the rigid boundary conditions. The influence of the other
part of the bulk through the soft boundary conditions will also be
taken into account in section III.

The uniaxial loading of a perfect crystallite under the rigid
boundary conditions can be realised, for example, at the central
contact of roughnesses at rolling of a rail by wheels of a
rolling-stock. The constants of the potential (1) and mass of
particles in certain conditional system of units are chosen to be
equal, $E_{b}=0.20833 mJ_{c}$, $r_{0}=1 m_{c}$, and $m=0.01
kg_{c}$, and the time  step is $t=0.18s_{c}$. The period of small
vibrations of particles is calculated with the formula
$T=2\pi\sqrt{m/U_{ij}''}$, where $U_{ij}$ is the total potential
energy of $i,j$-particle in the field of the other particles. In
the cases of a) two molecular system, b) non-linear chain and c)
vibration of crystalline "planes" in a two-dimensional crystal,
the periods of vibrations are equal to $5.12s_{c}, 3.63s_{c}$ and
$2.55s_{c}$, respectively. The velocities of sound waves in the
low-frequency limit in the both last  cases are equal to
$1.22m_{c}/s_{c}$ and $1.5m_{c}/s_{c}$, respectively.

\begin{figure}[p]
  \includegraphics [width=5.7 in] {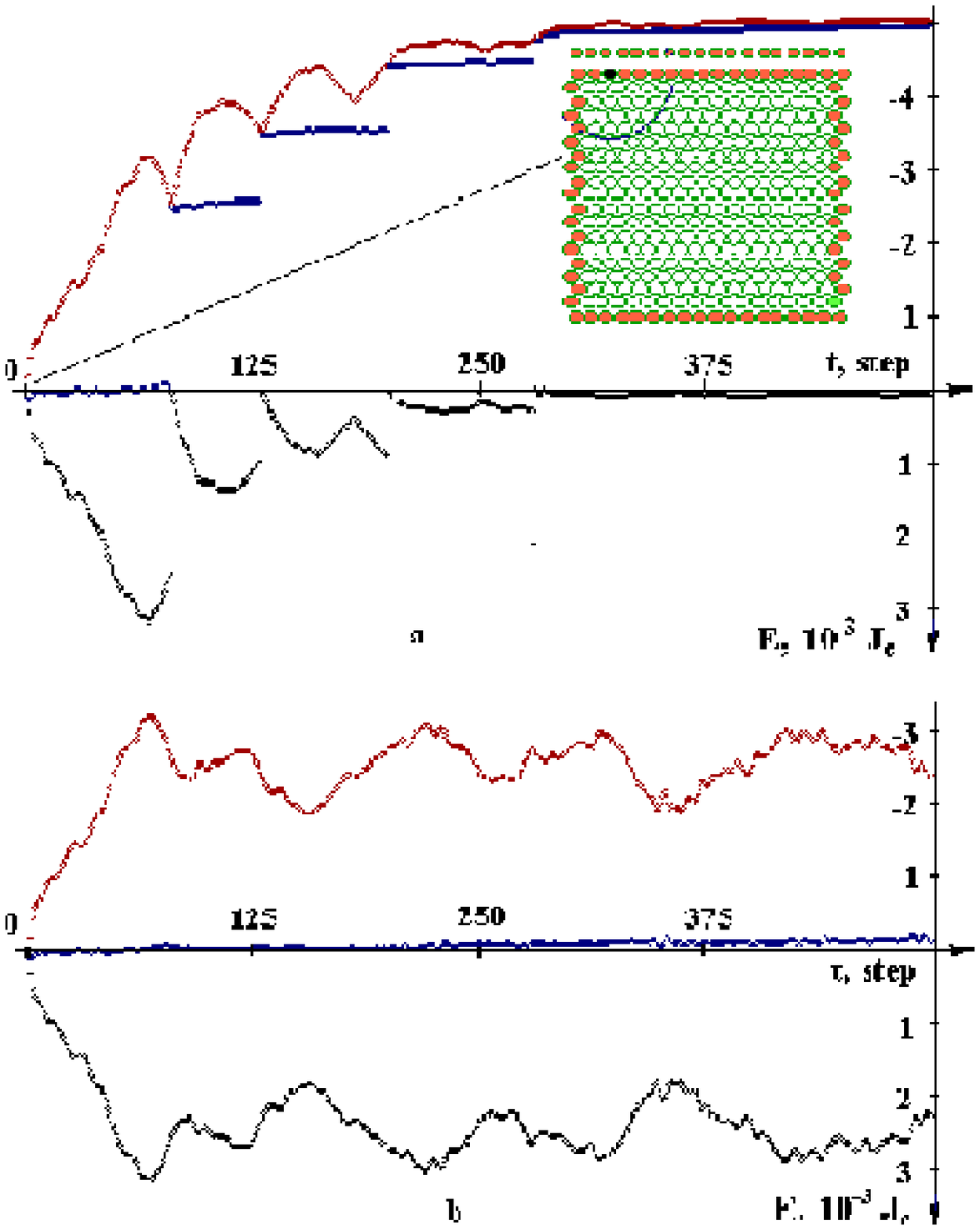}\\
  \caption{Changes of the general potential, kinetic and
internal energy of system on the initial stage:
  a - base experiment,
  b - experiment without cooling.}\label{f1}
\end{figure}

The crystallite was placed on a rigid motionless platform
consisting of atoms of the same sort, as the sample (the lowermost
atomic layer). The same rigid atomic layer goes from above
downwards with the constant velocity of $2.778 10^4 m_{c}/s_{c}$
(the uppermost atomic layer). The lateral sides of the sample are
free. The initial atomic configuration is presented by the
hexagonal lattice with the interparticle distance $r_{0}=1m_{c}$.
The initial configuration of the rigid boundaries and the system
as a whole is shown on the insert in fig.\ref{f1}a. In the initial
configuration between the rigid top side and the sample there is a
small gap.

In the base experiment, in the beginning and through every $29900$
time steps were executed the deep cooling of the sample by
five-multiple zeroing of kinetic energy of the atoms on $80, 130,
200, 280$ and $284$-th time steps. Such cooling liquidates surplus
of heat arising because of higher strain rate, than in a real
experiment. The additional details of the description of the base
experiment may be found in ref. \cite{Metlov}. Here for comparison
the computer experiments were distinguished from the base one a)
by the absence of periodic cooling of the sample through $29900$
time steps (cooling only at the initial stage) b) the absence of
cooling.

Change of different kinds of energy during the initial stage of
strain is given in the fig.\ref{f1}. At the expense of attraction
to the rigid boundary separated from the sample by a small gap the
sample in the beginning is slightly stretched, thus, in a
consequence of this, the low-frequency oscillations covering all
the system are arisen (Fig.\ref{f1}b). The period of the
low-frequency oscillations equals approximately $21s_{c}$. The
small change of the internal energy is negative. It testifies that
at this stage the system itself performs work on the moving of the
rigid boundary. In due course, the low-frequency oscillations are
dissipated, and their energy completely transforms to heat. In
this limit, the double average kinetic energy per one degree of
freedom, to within Boltzmann's constant, is a measure of
temperature of a sample \cite{wert64}. The energy removal at
cooling reduces its total potential and internal energy
(Fig.\ref{f1}a). In the both cases under consideration the
potential energy decreases in comparison with its initial value,
however, in the experiment with cooling it decreases by a larger
magnitude $(4,838 mJ_{c})$.

In the regarded cases the overall picture of crystallite strain at
the uniaxial compression is similar. After the stage of elastic
strain in the left top and right bottom corners of the sample two
dislocations are almost simultaneously formed. According to the
introduced in \cite{Metlov} definitions, they can be considered as
"elementary particles" of a kind $\pi_{2}^{+}$ and $\pi_{2}^{-}$.
At further strain the dislocations move along their own planes of
sliding up to those, do not achieve the rigid boundaries, where
are stopped. The stop of the dislocations results in an additional
elastic energy concentration around them. At some time the
puncture of material between the dislocations similarly to a
lightning stroke or electrical discharge occurs. The regrouping of
atoms owing to the puncture results in turning the both
dislocations by $60^{\circ}$ and their planes of sliding proved to
be parallel to the rigid boundaries. The last circumstance
promotes their further pressing out of the sample. As a result,
the sample completely restores its perfect crystal structure. With
growth of the strain the picture qualitatively repeats more then
once with the same script, - birth of dislocations, their
movement, turning out and leaving the sample, and complete renewal
of the crystal structure.

The irreversibility of the process is expressed by a change of the
lateral surfaces relief of the sample after each exit of the
dislocations, and with reduction of number of the layers in the
vertical direction just one. In the end, the work of external
forces is spent for increasing the bulk elastic energy, the
potential energy of the curved lateral surfaces, and for
increasing the energy of the thermal movement. Actually in this
picture dislocations are an intermediate link of transformation
(relaxation) of energy or an intermediate state. With temperature
increase the birth of dislocations is facilitated, the interval
between cycles is reduced and, at last, they merge in one
continuous process.

\begin{figure}[p]
  \includegraphics [width=7 in] {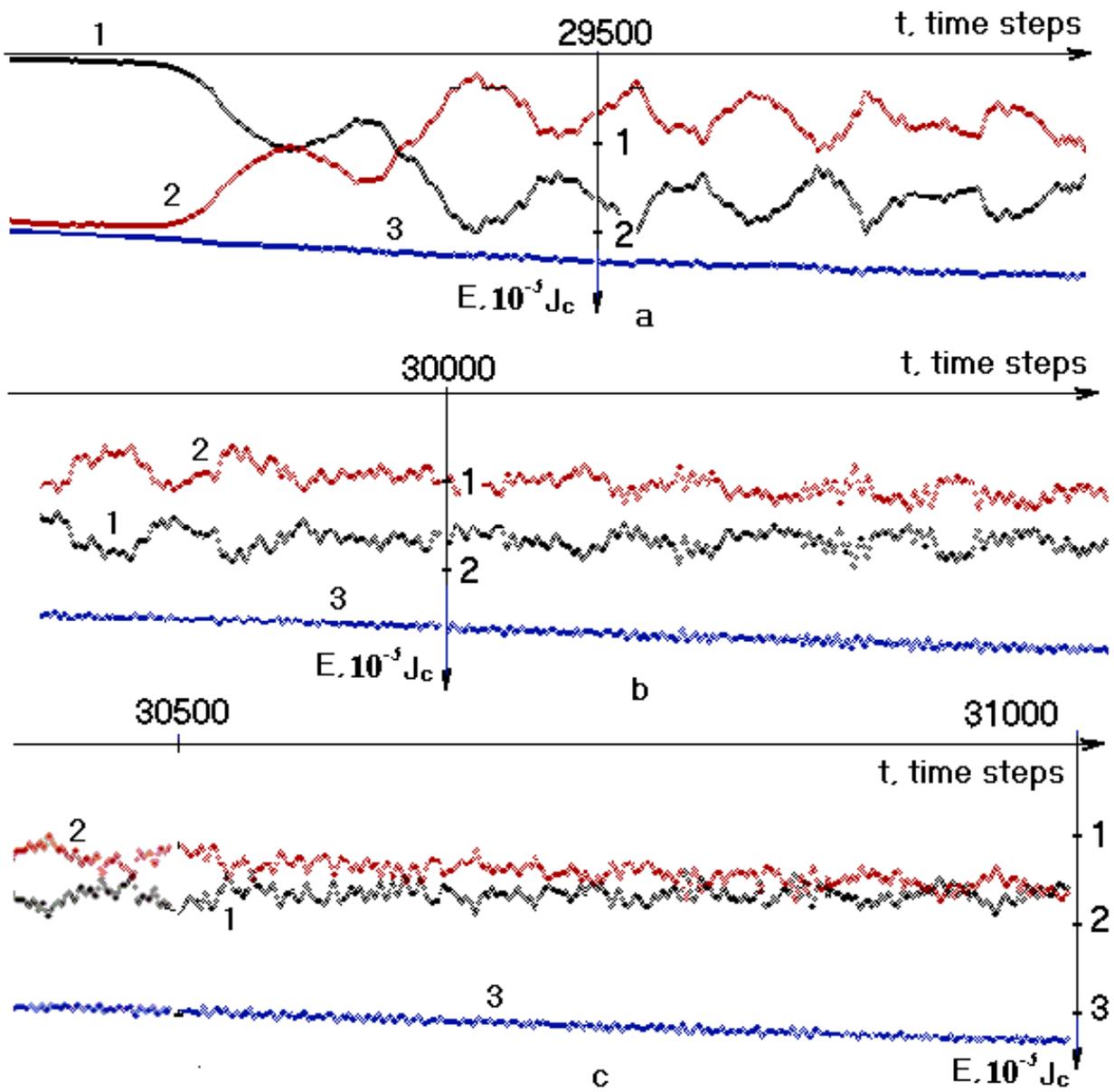}\\
  \caption{Elastic vibrations of nanobulk at stripping of the
first dislocations and their change during transformation into the
thermal fluctuations (cooling only on the initial
stage)}\label{f2}
\end{figure}

At different stages of system evolution features of energy
transformations are of special interest. In the experiments with
initial cooling, the kinetic energy is close to zero, and the
potential and internal energies are approximately the same down to
the moment of birth of a dislocation pair (Fig.\ref{f2}). The
dislocation pair birth on a background of slow growth of the
internal energy of the system at the expense of work of the rigid
boundaries is accompanied by conversion of a part of the potential
energy (elastic pressure) into the kinetic energy. The transition
is accompanied by excitation of elastic low-frequency vibrations
of the resonant character, with frequency gradually growing in due
course, and the amplitude falling. During about a two tens periods
of the low-frequency vibrations the latter calm down, and
completely pass into the thermal fluctuations. On record of the
transition the potential energy decreases by $1.164 mJ_{c}$.

\begin{figure}[p]
  \includegraphics [width=7 in] {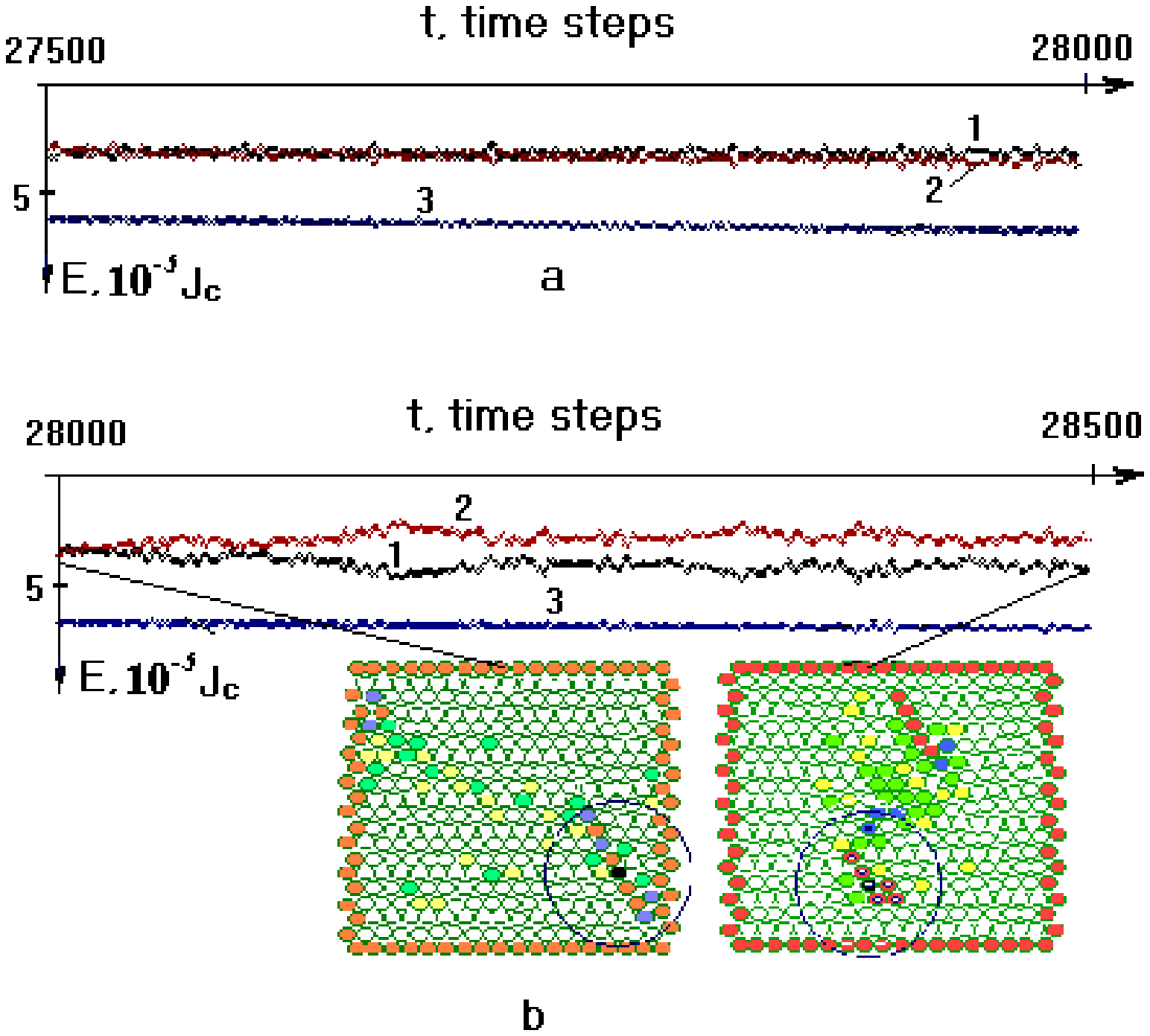}\\
  \caption{Elastic vibrations of nanobulk at stripping of first
dislocations and their change during transformation into the
thermal fluctuations (the cooling is absent at all). On inserts
the
  Yellow circles present the atoms with potential energy $0.02mJ_{c}<U<0.04mJ_{c}$,
  green circles - with $0.04mJ_{c}<U<0.1mJ_{c}$,
  blue circles - with $0.1mJ_{c}<U<0.15mJ_{c}$,
  red circles - $0.15mJ_{c}<U<0.38mJ_{c}$,.
  Potential energy of the particles on free lateral surfaces is equal $0.6mJ_{c}$.}
\label{f3}
\end{figure}

Without the prior cooling the described process gets some other
character (Fig.\ref{f3}). Higher initial value of the internal
energy leads to a more earlier birth of dislocation pair,
approximately, by $1000$ time steps. "Premature birth" results in
the fact that the dislocations are languid and inactive, as the
overall store of the elastic energy for setting them in motion, is
not so great yet. The dislocations slowly enough during $300-350$
time steps advance in directions of the rigid boundaries. The jump
of the potential energy during the transition is equal to $0.537
mJ_{c}$, that it is much less, than with cooling.

At turn of the dislocations, similar transformations of energy are
taking place too. In these cases the jumps of the potential energy
are equal, accordingly, to $1.041 mJ_{c}$ with the cooling (see
fig.1 in ref. \cite{Metlov}) and $2.608 mJ_{c}$ with no cooling.

\begin{figure}[p]
  \includegraphics [width=7 in] {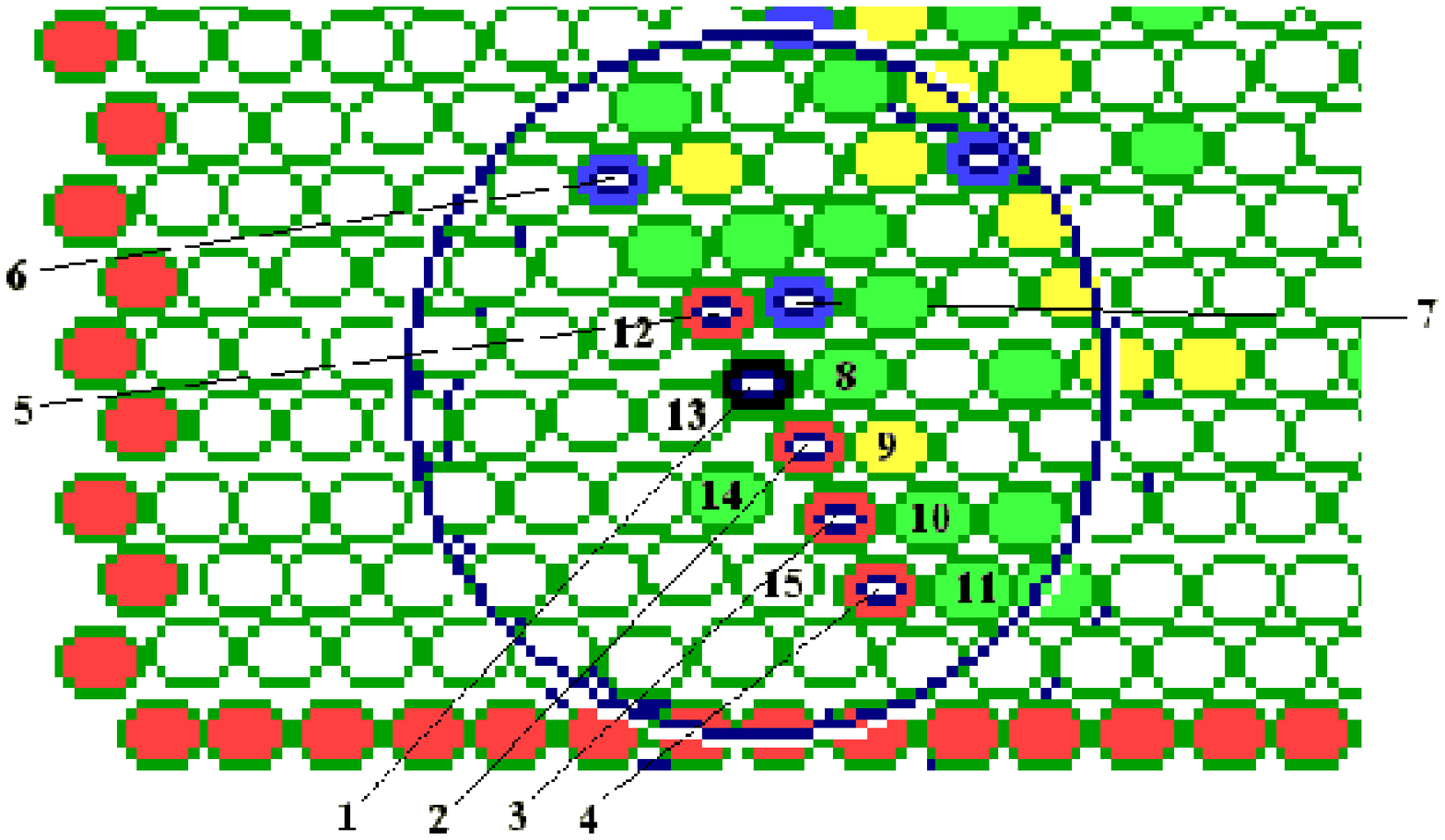}\\
  \caption{Values of the potential energy of particles in a
vicinity of the nucleus of dislocations (in mJc): 1 - 0.429; 2 -
0.403; 3 - 0.268; 4 - 0.202; 5 - 0.19; 6 - 0.129; 7 - 0.101; 8 -
0.058; 9 - 0.025, 10 - 0.063, 11 - 0.067, 12 - 0.019; 13 - 0.012;
14 - 0.041; 15 - 0.002. For a zero level the average potential
energy on one particle in the sample is accepted.} \label{f4}
\end{figure}

Example of potential energy distribution around a dislocation is
presented in fig.\ref{f4}. The atoms of the largest energy are
placed on the "compressed" chain. Next series of atoms ranged in
energy is placed behind the "compressed" chain, and only the next
series - on the "stretched" chain.

\section{Uniaxial compression of sample under soft boundary conditions}

The rigid boundary conditions as though completely, both in
thermodynamic and in mechanical sense, isolate the allocated bulk
from other part of a crystal. If this is possible to assume from
the point of view of thermodynamic isolation, as the heat
equilibrium before all is established in a small bulk, from the
point of view of far-acting mechanical fields it is not always
justified. The soft boundary conditions are reached at the expense
of introduction of the periodic boundary conditions along the
loading axis. In this connection there arises a question, whether
- will be and as far as to differ character of dislocation
behaviour from one under the rigid boundary conditions?

Strain is given by change of the periodicity size with the same
velocity, as movement of the rigid boundary in the previous
experiments. First dislocations in system consisting of 30*30
particles with binding energy $E_{b}=0.2083 mJ_{c}$ occur at the
$46839$-th time step. According to the accepted definition (4) in
ref. \cite{Metlov} they are dislocations of a type $\pi_{3}^{-}$,
$\pi_{2}^{-}$, $\pi_{3}^{+}$, $\pi_{2}^{+}$.

\begin{figure}[p]
  \includegraphics [width=7 in] {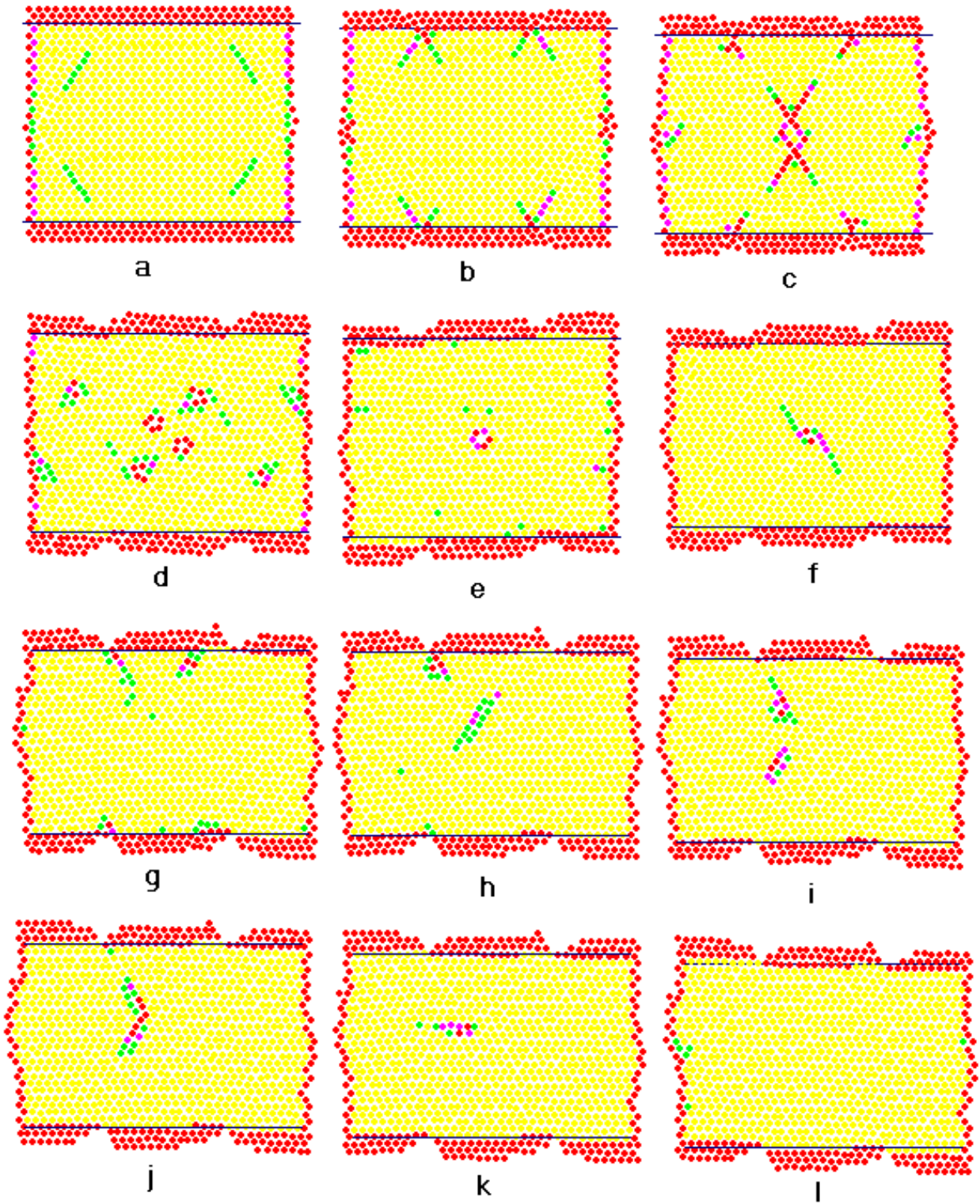}\\
  \caption{Evolution of structure of crystallite at soft strain.
The images correspond to the following time steps: a - 46840, b -
46850, c - 46880, d - 46990, e - 47240, f - 83152, g - 83630, h -
83715, i - 83816, j - 84005, k - 84040, l - 117000; a-e - first
cycle of not elastic strain, f-k - second cycle of not elastic
strain, l - beginning of the third cycle of not elastic
strain.}\label{f5}
\end{figure}

In the given series of experiments the dislocations arise not at
edges of up and down boundaries, as in the case of the rigid
loading, but at the central parts of free lateral sides of the
sample (Fig.\ref{f5}). They move in the directions of the
horizontal boundaries and pass through them (Fig.\ref{f5}b). As in
such system the stoppers for movement of dislocations, playing the
important role in the previous example, are absent, the
dislocations continue to move at achievement of the horizontal
boundaries. As a result, they meet at the centre of the sample
(Fig.\ref{f5}c). Further the picture becomes complicated and in
the sample 5, 6 dislocations are observed simultaneously, and,
they transit into the structural state "with micropore"
(Fig.\ref{f5}d). At $47240$-th time step a part of dislocations
come out on the free lateral surfaces, derivating roughnesses of
its relief, the others annihilate among themselves, derivating the
vacancy at the centre of the sample, and in the system the perfect
crystal structure is restored. The described events occur during
$400$ time steps, that is in current the short interval of time in
comparison with the previous phase of elastic strain ($46800$ time
steps). Furthermore, the system calms down for a long time, and
during $36000$ time steps the dislocations are absent at all.
During this time the strain again proceeds in the elastic manner.

At the $83152$-th time step the vacancy at the centre of the
sample breaks up to the pair of dislocations $\pi_{3}^{+}$ and
$\pi_{2}^{-}$ (Fig.\ref{f5}f), which move in different directions
(Fig.\ref{f5}g). The whole series of births and annihilations of
dislocations begins with this decay which is not given completely
in the fig.\ref{f5}. From this series the fragments, illustrating
the merge of two dislocations and the birth of third ones
(Fig.\ref{f5}) are only given. At the $83630$-th time step from
the top boundary begin to move dislocations $\pi_{2}^{-}$ and
$\pi_{3}^{+}$. At first the dislocation $\pi_{3}^{+}$ outstrips
the dislocation $\pi_{2}^{-}$, crossing the sliding plane of the
last-mentioned before arrival of one to this point (Fig.\ref{f5}i,
j). At that instant, when the dislocation $\pi_{2}^{-}$ reaches
the point of crossing the sliding planes (Fig.\ref{f5}i), it
begins to attract the first one. As a result of this attraction
the first dislocation comes back along its own sliding plane
(Fig.\ref{f5}j), both dislocations merge and form a new one
(Fig.\ref{f5}l). The latter is pressed out from the sample in
parallel to its horizontal boundaries. As a result, in the system
perfect crystal structure, already without vacancy is again
restored. The following stage of inelastic strain begins
approximately through $28000$ time steps (Fig.\ref{f5}m).

Thus, it is possible to ascertain, that the general feature, - the
cyclic change of elastic and inelastic stages of strain, marked at
pressing by the rigid boundaries, is kept in the case of the soft
boundaries too.

\section{Cutting sample by three-nuclear knife}

In the previous experiments on a nanoscopic level an analogue of a
homogeneous and non-hydrostatic state was realised. The influence
of a non-uniform loading on character of generation and behaviour
of dislocations is of large interest too. This problem was
investigated in the next series of computer experiments simulating
the cutting of the same sample by a rigid three-atomic knife
\cite{Metlov}. In the moving downwards rigid top boundary the
three central atoms are only left. The binding energy of particles
in these experiments is accepted to be equal to $0.0416 mJ_{c}$,
other parameters are same, as in section 1.

The initial elastic stage of strain ends, when in the region of
three-nuclear indentor first pair of dislocations is born, the
sliding planes of which form an angle of $60^{\circ}$. Then they
move to the bottom rigid boundary, are turned and leave the
sample. However they leave the sample not with parallel rigid side
as in a case of uniaxial compression, but mainly at an angle of
$60^{\circ}$ to it, being directed to the left or right top
corners of the sample (see fig. 2 in the ref. \cite{Metlov}).

However at general similarity of the script the character of
dislocations is different, - they pass in some new structure
modification with micropore at the dislocation nucleus
\cite{Metlov}. Owing to micropore a dislocation becomes more
localized, than in the usual allocated state. The atomic planes in
area divided by a normal to the plane of sliding, diverge at a
more larger angles than in the previous example. Such dislocation
can be considered as a possible element of large-angle grain
boundaries, which at small quantity of these elements can result
in the formation of curved boundaries and, as a consequence, in
the formation of a fine grain.

Thus, in the previous series of experiments has been shown that
dislocations, having the same Burgers vector, can be, at least, in
two structure modifications or states, - in allocated (basic)
states and ones with a micropore. A distinctive attribute of a
state with micropore is the presence on the lattice images of
typical pentagons. It is characteristic, that a "state with
micropore" occurs at non-uniform or intense loading more
preferably. Except for the mentioned above two structure
modifications there are also others \cite{Metlov}, which can have
the large importance in dynamics of IPD.

\section{Shear strain of sample by rigid shell}

A plenty of dislocations arises at straining a sample under the
shear loading. This loading is realised as follows: the sample
from 28*28 particles with binding energy of $0.0416 mJ_{c}$ is
placed in the rigid two-dimensional shell consisting of atoms of
the same type, as the sample. The shell is deformed through
movement of the top and bottom sides of the shell to itself with
constant velocity in mutually opposite directions.

At an initial stage of strain, dislocations $\pi_{3}^{-}$ and
$\pi_{2}^{-}$ are born (see classification and fig. 3 in ref.
\cite{Metlov}). Then the first of them decays on two new
dislocations according to the equation of reaction:
\begin{equation}\label{a2}
    \pi_{3}^{-}+ energy\rightarrow\pi_{1}^{-}+ \pi_{2}^{-}.
\end{equation}
(In formula (7) of ref. \cite{Metlov} in recording of this
reaction there was an annoying misprint).

Strain under the uniaxial loading leads to a repeated recurrence
of the same common script of system behaviour, - birth of
dislocations, them turning out, restoration of perfect crystal
structure etc. At strain of a sample under the shear loading the
system evolution proceeds under essentially different script. In
view of special kinematics of such strain the restoration of
perfect structure is possible only after a turn of some
"macroscopic" bulk as whole through a finite discrete angle being
an element of crystal symmetry. For two-dimensional problem it is
the angle of $60^{\circ}$. For realization of such a turn it is
impossible to do with a sequence of births and annihilations of
one or two dislocations only. The system would have to accumulate
set of dislocations, and, as a consequence to form new boundaries
and fine grains to restore the symmetry even if in a local region.
This feature determines the shear loading of strain as the
effective tool of material structure transformation. And in most
cases dislocations are not in the basic structural state, and in a
"state with micropore", which is testified by the typical pentagon
fragments on the images of a lattice. Owing to "pentagons" the
strongly bent boundaries of fine grains are formed.

\section{Summary}

    Thus, the kinetics of dislocations is observed in different
numerical experiments. It is established that the dislocations may
have at least two different structural states. The first of them
has whole stretched chains, the second has micropores. The
dislocations with micropores, as a rule, are of pentagon form.
Owing to micropore a dislocation becomes more localised, than in
the usual allocated state. The atomic planes in area divided by a
normal to the plane of sliding, diverge at more larger angles.
Such dislocation can be considered as a possible element of
large-angle grain boundaries, which at small quantity of these
elements can result in the formation of curved boundaries and, as
a consequence, in the formation of a fine grain. Strain under the
uniaxial or "knife" loadings lead to repeated recurrence of the
same common script of system behaviour, - birth of dislocations,
their turning out, restoration of perfect crystal structure etc.
Alternation of elastic and inelastic strain stages leads one to
the assumption that the studied sample can be alternatively as
elastic or non-elastic element of more complex bodies such as Foit
or Maxwell ones.

At straining a sample under the shear loading the system evolution
proceeds under essentially other script. In view of special
kinematics of such strain the restoration of perfect structure is
possible only after turning some "macroscopic" bulk as whole
through a finite discrete angle being an element of crystal
symmetry. For realization of such turn it is impossible to do with
sequence of births and annihilations of one or two dislocations
only. The system would have to accumulate set of dislocations,
and, as a consequence to form new boundaries and fine grains to
restore the symmetry even if in a local region. This feature
determines the shear loading of strain as an effective tool of
material structure transformation.

\bibliography{MD2}

\begin{thebibliography}{5}
\expandafter\ifx\csname natexlab\endcsname\relax\def\natexlab#1{#1}\fi
\expandafter\ifx\csname bibnamefont\endcsname\relax
  \def\bibnamefont#1{#1}\fi
\expandafter\ifx\csname bibfnamefont\endcsname\relax
  \def\bibfnamefont#1{#1}\fi
\expandafter\ifx\csname citenamefont\endcsname\relax
  \def\citenamefont#1{#1}\fi
\expandafter\ifx\csname url\endcsname\relax
  \def\url#1{\texttt{#1}}\fi
\expandafter\ifx\csname urlprefix\endcsname\relax\def\urlprefix{URL }\fi
\providecommand{\bibinfo}[2]{#2}
\providecommand{\eprint}[2][]{\url{#2}}

\bibitem[{\citenamefont{Pashinskaya et~al.}(2002)\citenamefont{Pashinskaya,
  Metlov, Varyukhin, and Morozov}}]{PMVM02}
\bibinfo{author}{\bibfnamefont{E.~G.} \bibnamefont{Pashinskaya}},
  \bibinfo{author}{\bibfnamefont{L.~S.} \bibnamefont{Metlov}},
  \bibinfo{author}{\bibfnamefont{V.~N.} \bibnamefont{Varyukhin}},
  \bibnamefont{and} \bibinfo{author}{\bibfnamefont{A.~F.}
  \bibnamefont{Morozov}}, \bibinfo{journal}{Proceeding of the V International
  Conference Metallurgy, Refractories and Environment, Stara Lesna, High
  Tatras, Slovakia, May 13-16}  (\bibinfo{year}{2002}).

\bibitem[{\citenamefont{Valiev and Alexandrov}(2000)}]{Vali00}
\bibinfo{author}{\bibfnamefont{R.~Z.} \bibnamefont{Valiev}} \bibnamefont{and}
  \bibinfo{author}{\bibfnamefont{I.~V.} \bibnamefont{Alexandrov}},
  \emph{\bibinfo{title}{Nano-strucrure materials resulting with intense plastic
  strains}} (\bibinfo{publisher}{Logos}, \bibinfo{address}{Moscow, Russia},
  \bibinfo{year}{2000}), \bibinfo{note}{in Russian}.

\bibitem[{\citenamefont{Korshunov et~al.}(2001)\citenamefont{Korshunov, Efros,
  Chernenko, and Goychenberg}}]{KEChG01}
\bibinfo{author}{\bibfnamefont{L.~G.} \bibnamefont{Korshunov}},
  \bibinfo{author}{\bibfnamefont{B.~M.} \bibnamefont{Efros}},
  \bibinfo{author}{\bibfnamefont{N.~L.} \bibnamefont{Chernenko}},
  \bibnamefont{and} \bibinfo{author}{\bibfnamefont{Y.~N.}
  \bibnamefont{Goychenberg}}, \bibinfo{journal}{Phizika i Tehnika visokih
  davleniy} \textbf{\bibinfo{volume}{11}}, \bibinfo{pages}{75}
  (\bibinfo{year}{2001}), \bibinfo{note}{in Russian}.

\bibitem[{\citenamefont{Metlov}(2002)}]{Metlov}
\bibinfo{author}{\bibfnamefont{L.~S.} \bibnamefont{Metlov}},
  \bibinfo{journal}{http://arxiv.org/abs/cond-mat/02100486}
  p.~\bibinfo{pages}{1} (\bibinfo{year}{2002}), \bibinfo{note}{print 23 Oct
  2002}.

\bibitem[{\citenamefont{Wert and Thomson}(1964)}]{wert64}
\bibinfo{author}{\bibfnamefont{C.~A.} \bibnamefont{Wert}} \bibnamefont{and}
  \bibinfo{author}{\bibfnamefont{R.~M.} \bibnamefont{Thomson}},
  \emph{\bibinfo{title}{Physics of Solids}} (\bibinfo{publisher}{McGraw-Hill,
  Book Company}, \bibinfo{address}{New-York - San Francisco - Toronto -
  London}, \bibinfo{year}{1964}).

\end{thebibliography}

\end{document}